\documentclass[aps,prb,floatfix,epsfig,preprint,showpacs,preprintnumbers]{revtex4}

\usepackage{amssymb}
\usepackage{graphicx}
\usepackage{amsmath}

\begin{document}

\title{Electronic structure of Pu and Am metals by self consistent
relativistic GW method}
\author{Andrey Kutepov$^{\dag }$, Kristjan Haule$^{\dag }$, Sergey Y.
Savrasov$^{\ast }$, Gabriel Kotliar$^{\dag }$}
\affiliation{$^{\dag }$Department of Physics,Rutgers University, Piscataway, NJ 08856}
\affiliation{$^{\ast }$Department of Physics, University of California, Davis, CA 95616}

\begin{abstract}
We present the results of calculations for Pu and Am performed using an
implementation of self--consistent relativistic GW method. The key feature
of our scheme is to evaluate polarizability and self--energy in real space
and Matsubara's time. We compare our GW results with the calculations using
local density (LDA) and quasiparticle (QP) approximations and also with
scalar--relativistic calculations. By comparing our calculated electronic
structures with experimental data, we highlight the importance of both
relativistic effects and effects of self--consistency in this GW calculation.
\end{abstract}

\pacs{71.27.+a, 71.15.Rf, 71.20.Gj}
\maketitle

\section{Introduction}

\label{intro}

During the last two decades we have been witnessing a surge of activity in
many--body--theory based methodologies applied to condensed matter physics.
Here we are concerned with one of them, the Hedin's GW method \cite%
{prb_139_A796}. This particular approach has not only been applied to many
different materials but it has also been formally developed with an intent
to enhance its own applicability or to diagrammatically extend it.

First applications of GW were of "one--shot" type when one starts with local
density approximation (LDA) to get one--electron eigenstates and construct
the corresponding Green's function which is then used as an input to perform
only one GW iteration. Commonly, such an approach is called G$_{0}$W$_{0}$.
It usually improves LDA band gaps in semicoductors\cite{rpp_61_237} but has
an obvious drawback because the absence of self--consistency makes it
depending on input and not conserving.\cite{pr_127_1391,prb_64_235106}

To make the approach independent on the input, the quasi--particle
self--consistent GW method (QSGW) was introduced a few years ago.\cite%
{prl_96_226402,prb_76_165106} In this method the Green's function is found
self--consistently with approximate Hermitian form of self--energy which is
constructed to minimize the perturbation while keeping a quasi--particle
picture. The approach was successfully applied to a wide class of materials
including simple metals, semiconductors, wide band gap insulators,
transition metals, transition metal oxides, magnetic insulators, and rare
earth compounds. First calculations for actinide metals using this
approximation and neglecting spin--orbit interaction have also been reported.%
\cite{prb_78_081101,philmag_89_1801} Recently a scheme based on L\"{o}wdin's
orthogonalization was proposed\cite{prb_80_235128} which removes an
ambiguity in the construction of the effective self--energy in QSGW. The
method has also been extended to treat finite temperatures\cite%
{prb_74_033101} and to calculate spin wave dispersions\cite{jcm_20_295214}.
However, similar to the "one-shot" variants of GW, QSGW method is not $\Phi $%
-derivable\cite{pr_127_1391}, and, as a consequence, it is not conserving.
This, for example, results in difficulties to calculate total energy.

Applications of fully self--consistent GW schemes are not numerous. They
have been applied for weakly correlated solids\cite%
{prl_81_1662,prl_89_126402,prl_96_226403,prb_80_041103} and for free atoms
and molecules\cite{epl_76_298, jchp_130_114105}. General conclusion seems to
be that for weakly correlated simple solids full self--consistency
deteriorates spectra as compared to "one--shot" or QSGW approximations but
improves total energies. For free atoms the conclusion clearly favors fully
self--consistent calculations. Based on these facts one can expect that in
solids the spectra obtained by fully self--consistent GW might be
competitive with spectra from QSGW if the corresponding physics is local
enough, i.e. similar to free atoms. Besides, the fully self--consistent GW
is $\Phi $--derivable and so it is conserving. Also, it is important to
mention the works aimed to enhance the accuracy of GW based schemes, their
robustness, performance, and convergency issues \cite%
{prb_74_045102,prb_74_245125,cpc_117_211,prb_74_035101,prb_75_235102,
prb_74_045104,prb_81_125102}.

Another very active field related to the GW method is its diagrammatic
extensions. We mention here the approaches which use LDA--based vertex
correction\cite{prb_49_8024,prl_99_246403,prb_54_7758,prb_81_085213}, the
approaches which use direct diagrammatic representation for the vertex\cite%
{prb_40_11659,prb_43_1851,prb_54_2374,prb_57_11962,prl_80_1702}, and the
approach which combines GW and dynamical mean field theory (GW+DMFT)\cite%
{prl_90_086402,jcm_17_7573}. Hedin's equations and correspondingly the GW
method have also been formally extended to spin--dependent interactions\cite%
{prl_100_116402,jcm_21_064232}, to treat the electrons residing in a
subspace of the full Hilbert space\cite{prl_102_176402}, and onto the
Keldysh time--loop contour\cite{arx.1106.1094}.

Very recently the importance of spin--orbit interaction was highlighted for
the elements with large atomic numbers and it was perturbatively included in
"one--shot" GW calculations for $Hg$ chalcogenides\cite{prb_84_085144}. In
this work we generalize the GW method to solve equations explicitly based on
4--component Dirac's theory, which is important to get meaningful results
for such elements as actinides. This fact together with uncertainty in
respect to what kind of self--consistency\ is better to use for actinides
defines the scope of the present work in which we apply self--consistent GW
method based on Dirac equation to study the electronic structure of
Plutonium and Americium metals.

These two metals (especially Pu) have been a subject of intensive studies
during last two decades. From theoretical point of view the best
understanding\cite{nature_410_793,prb_75_235107,nature_446_513,
prl_101_056403,prl_101_126403,prb_82_085117} was achieved using a
combination of LDA and dynamical mean field theory\cite{rmp_68_13} (DMFT)
known as LDA+DMFT method. LDA+DMFT calculations have resolved the puzzle of
false magnetism in Pu and Am metals which appears in density--functional
based calculations\cite{epl_55_525,jcm_15_2607,jmmm_272_329,prb_77_085101}
but contradicts\ with the experiment \cite{prb_72_054416,arxiv_0508694}.

However, there is a problem with LDA+DMFT type of calculations as the
approach is not parameter--free and requires the input matrix of on--site
Hubbard interactions. On top of that there is an uncertainty with double
counting correlation effects that are present both in LDA and DMFT theories.
Therefore there is a significant interest to develop diagramatically based
approaches such as GW and its extensions that offer the possibility to
overcome both problems. In respect to Plutonium, our study can be considered
as the extension of previous work by Chantis et al\cite{philmag_89_1801} who
have studied this metal with QSGW without spin--orbit interaction and
concluded that correlation effects included in GW make the $f$--bands
narrower and decrease the crystal--field splittings as compared to the LDA
results. We extend the work [\onlinecite{philmag_89_1801}] in three ways: i)
include spin--orbit interaction by using Dirac form for kinetic energy
operator, ii) perform fully self--consistent GW calculation and compare it
with self--consistent quasi--particle (QP) and local density approximations,
and iii) apply self--consistent GW method to Am metal.

\section{Relativistic GW Method}

\label{RGW}

Althought, a truly relativistic treatment of the problem would require the
use of rather complicated equations of Quantum Electrodynamics, we use a
simplified approach. First, we neglect relativistic retardation effects in
the Coulomb interaction. In this case, Hedin's original derivation of his
famous system of equations\cite{prb_139_A796} still holds with the only
extension that all fermionic functions (Green's function and self energy)
become $4\times 4$ matrices for every pair of space coordinates. Also, in
order to perform self--consistent GW calculation we need only scalar parts
of the bosonic functions (polarizability $P$ and screened interaction $W$)
in the space of products of bi--spinors which is similar to the
non--relativistic theory with collinear spin structures. So, in our method
which is described below only the fermionic functions (Green's function and
self--energy) have bi--spinor arguments. Second, we exclude positron states
and represent the coordinate dependence of Green's function in terms of
electron states only

\begin{align}\label{G_band_k}
G(\alpha \mathbf{r},\alpha ^{\prime }\mathbf{r}^{\prime };\tau )=\frac{1}{N_{%
\mathbf{k}}}\sum_{\mathbf{k}}\sum_{\lambda \lambda ^{\prime }}\Psi _{\lambda
}^{\mathbf{k}}(\alpha \mathbf{r})G_{\lambda \lambda ^{\prime }}^{\mathbf{k}%
}(\tau )\Psi _{\lambda ^{\prime }}^{\dag ,\mathbf{k}}(\alpha ^{\prime }%
\mathbf{r}^{\prime }),
\end{align}%
where $\mathbf{k}$ runs over Brillouin zone, $N_{\mathbf{k}}$ is the number
of $\mathbf{k}$-points, indexes ($\lambda ,\lambda ^{\prime }$) denote the
electronic Bloch band states, as obtained from relativistic LDA\cite%
{jcm_15_2607} or Hartree--Fock (HF) problem, ($\alpha ,\alpha ^{\prime }$)
are the bi--spinor arguments, and $\tau $ is Matsubara's time. Thus, in the
coordinate--space representation, Green's function is generally $4\times 4$
matrix for every $\mathbf{r},\mathbf{r}^{\prime }$ pair.

Inside the muffin--tin (MT) spheres the Bloch states can conveniently be
represented as linear combinations of 4--component solutions $\varphi _{LE}^{%
\mathbf{t}}(\alpha \mathbf{r})$ of radial Dirac equation taken with the
spherical symmetric part of Hamiltonian inside the sphere

\begin{align}\label{bands_mt}
\Psi _{\lambda }^{\mathbf{k}}(\alpha \mathbf{r})|_{\mathbf{t}}=\sum_{LE}Z_{%
\mathbf{t}LE}^{\mathbf{k}\lambda }\varphi _{LE}^{\mathbf{t}}(\alpha \mathbf{r%
}),
\end{align}%
where $\mathbf{t}$ is the specific atom in the unit cell, $L$ combines all
spin--angular quantum numbers, and index $E$ differs between $\varphi $, $%
\dot{\varphi}$, and local orbitals. Coefficients $Z_{\mathbf{t}LE}^{\mathbf{k%
}\lambda }$ ensure the smooth mapping between the muffin--tin spheres and
the interstitial region as it is standardly done in the linear agumented
plane wave (LAPW) method.

In the interstitial region we neglect by relativistic effects, i.e. we
assume the small components to be zero and represent the large components of
Bloch states as linear combinations of two--component spinors

\begin{align}\label{bands_int}
\Psi _{\lambda }^{\mathbf{k}}(\alpha \mathbf{r})|_{Int}=\frac{1}{\sqrt{%
\Omega _{0}}}\sum_{\mathbf{G}s}A_{\mathbf{G}s}^{\mathbf{k}\lambda
}u_{s}(\alpha )e^{i(\mathbf{k}+\mathbf{G})\mathbf{r}},
\end{align}%
where $\mathbf{G}$ runs over reciprocal lattice vectors; s is spin index, $%
\Omega _{0}$ is the unit cell volume, $u_{s}(\alpha )$ is a two--component
spin function, and $A_{\mathbf{G}s}^{\mathbf{k}\lambda }$ are the
variational coefficients in the LDA eigenvalue problem. We keep the same
bi--spinor argument $\alpha $ here with understanding that two of four
components at every $\mathbf{r}$ point in the interstitial region are
approximated to zero. Such an approximation greatly reduces computational
time for GW but is still well justified because relativistic effects are
mostly confined near the nuclei. We have checked the quality of this
approximation by performing LDA calculations with and without relativistic
treatment of the interstitial region, and the differences appear to be very
small.

In our implementation of the GW method we have taken an advantage of the
well known fact that polarizability and self--energy in Hedin's GW system of
equations \cite{prb_139_A796} are most easily evaluated in $(\mathbf{r};\tau
)$--representation while the equations for Green's function and screened
Coulomb interaction are most easily solved in $(\mathbf{k};\omega /\nu )$%
--representation where $\omega /\nu $ denote fermionic/bosonic
Matsubara's frequencies. So, in our approach we switch from one
representation to another using Fast Fourier Transform (FFT)
algorithm whenever needed.

Below we give the most important formulae as they appear in the
course of one loop of the self--consistency. The expressions
(\ref{bands_mt}) and (\ref{bands_int}) allow us to express $G(\alpha
\mathbf{r},\alpha ^{\prime } \mathbf{r}^{\prime };\tau )$ for both
$\mathbf{r}$ and $\mathbf{r}^{\prime }$ being inside the MT spheres
as follows (due to the symmetry of the solid we can restrict
$\mathbf{r}$ to be inside the unit cell with $\mathbf{R}=0$
whereas $\mathbf{r}^{\prime }$ may be inside the unit cell with $\mathbf{R}%
^{\prime }\neq 0$).

\begin{align}\label{G_mt_mt0}
G_{\mathbf{t}\alpha \mathbf{r};\mathbf{t}^{\prime }\alpha ^{\prime }\mathbf{r%
}^{\prime }}^{\mathbf{R}^{\prime }}(\tau )=\sum_{EL;E^{\prime }L^{\prime
}}\varphi _{EL}^{\mathbf{t}}(\alpha \mathbf{r})G_{\mathbf{t}EL;\mathbf{t}%
^{\prime }E^{\prime }L^{\prime }}^{\mathbf{R}^{\prime }}(\tau )\varphi
_{E^{\prime }L^{\prime }}^{\dagger ,\mathbf{t}^{\prime }}(\alpha ^{\prime }%
\mathbf{r}^{\prime }),
\end{align}%
Here $\mathbf{r}$ is inside of atom $\mathbf{t}$ in the central unit cell, $%
\mathbf{r}^{\prime }$ is inside of atom $\mathbf{t}^{\prime }$ in the unit
cell $\mathbf{R}^{\prime }$, and the number of different $\mathbf{R}^{\prime
}$ is exactly equal to the number of $\mathbf{k}$--points inside the
Brillouin zone.

In case when both $\mathbf{r}$ and $\mathbf{r}^{\prime }$ are in the
interstitial region we have three different representations for Green's
function: i) numerical values on regular mesh $G(\alpha \mathbf{r},\alpha
^{\prime }\mathbf{r}^{\prime };\tau )$, ii) band states representation $%
G_{\lambda \lambda ^{\prime }}^{\mathbf{k}}(\tau )$ which follows from (\ref%
{G_band_k}), and iii) representation in terms of plane waves

\begin{align}\label{G_int_int0}
G^{\mathbf{R}^{\prime }}& (\alpha \mathbf{r},\alpha ^{\prime }\mathbf{r}%
^{\prime };\tau )=\frac{1}{N_{\mathbf{k}}}\sum_{\mathbf{k}}e^{-i\mathbf{k}%
\mathbf{R}^{\prime }}  \notag  \\
& \times \sum_{s\mathbf{G};s^{\prime }\mathbf{G}^{\prime }}e^{i(\mathbf{k}+%
\mathbf{G})\mathbf{r}}u_{s}(\alpha )G_{s\mathbf{G};s^{\prime }\mathbf{G}%
^{\prime }}^{\mathbf{k}}(\tau )u_{s^{\prime }}^{\dagger }(\alpha ^{\prime})e^{-i(%
\mathbf{k}+\mathbf{G}^{\prime })\mathbf{r}^{\prime }}.
\end{align}
We can easily transform between the representations i) and iii) using FFT
while the representation ii) is connected to iii) by the formula (\ref%
{bands_int}). Finally, when one of the arguments (say $\mathbf{r}$) is
inside the MT space and another one belongs to the interstitial region the
representations for Green's function are obtained as obvious combinations of
the formulae above.

We begin our GW self--consistent cycle by transforming Green's function from
($\mathbf{k},\tau $)--representation to the real--space. Then we calculate
the polarizability in ($\mathbf{r},\tau $)-variables. For the $\mathbf{r},%
\mathbf{r}^{\prime }$ pair of indexes within the MT spheres we have the
following expression

\begin{align}\label{pol_mm_2}
& P_{\mathbf{t}Lk;\mathbf{t}^{\prime }L^{\prime }k^{\prime }}^{\mathbf{R}%
^{\prime }}(\tau )=  \notag \\
& -\sum_{E_{1}L_{1}}\sum_{E_{3}L_{3}}\sum_{\alpha }\langle M_{Lk}^{\mathbf{t}%
}\varphi _{E_{3}L_{3}}^{\mathbf{t}}(\alpha )|\varphi _{E_{1}L_{1}}^{\mathbf{t%
}}(\alpha )\rangle   \notag \\
& \times \sum_{E_{2}L_{2}}G_{\mathbf{t}E_{1}L_{1};\mathbf{t}^{\prime
}E_{2}L_{2}}^{\mathbf{R}^{\prime }}(\tau )\sum_{E_{4}L_{4}}G_{\mathbf{t}%
E_{3}L_{3};\mathbf{t}^{\prime }E_{4}L_{4}}^{\ast ,\mathbf{R}^{\prime
}}(\beta -\tau )  \notag \\
& \times \sum_{\alpha ^{\prime }}\langle \varphi _{E_{2}L_{2}}^{\mathbf{t}%
^{\prime
}}(\alpha')|\varphi_{E_{4}L_{4}}^{\mathbf{t}^{\prime}}(\alpha')M_{L^{\prime
}k^{\prime }}^{\mathbf{t}^{\prime }}\rangle ,
\end{align}%
where indexes $k$ and $k^{\prime }$ distinguish bosonic basis functions $M_{%
\mathbf{r}}^{\mathbf{t}Lk}$ (product basis functions which are scalars in
bi--spinor space) with the same angular symmetry and we have omitted
argument $\mathbf{r}$ of all functions in the integrands. For the
MT--interstitial and interstitial--interstitial combinations of $\mathbf{r},%
\mathbf{r}^{\prime }$ we obtain:

\begin{align}  \label{pol_mi_2}
P^{\mathbf{R}^{\prime }}_{\mathbf{t}Lk;\mathbf{r}^{\prime
}}(\tau)&=-\sum_{E_{1}L_{1}}\sum_{E_{2}L_{2}}\sum_{\alpha}\langle M^{\mathbf{%
t}}_{Lk}\varphi^{\mathbf{t}}_{E_{2}L_{2}}(\alpha)| \varphi^{\mathbf{t}%
}_{E_{1}L_{1}}(\alpha)\rangle  \notag \\
&\times\sum_{\alpha^{\prime }} G^{ \mathbf{R}^{\prime }}_{\mathbf{t}%
E_{1}L_{1};\alpha^{\prime }\mathbf{r}^{\prime }}(\tau)G^{*, \mathbf{R}%
^{\prime }}_{\mathbf{t}E_{2}L_{2};\alpha^{\prime }\mathbf{r}^{\prime
}}(\beta-\tau).
\end{align}

\begin{align}  \label{pol_ii_2}
P^{\mathbf{R}^{\prime }}_{\mathbf{rr}^{\prime
}}(\tau)=-\sum_{\alpha\alpha^{\prime }}G^{\mathbf{R}^{\prime }}_{\alpha%
\mathbf{r}\alpha^{\prime }\mathbf{r}^{\prime }}(\tau) G^{*,\mathbf{R}%
^{\prime }}_{\alpha\mathbf{r}\alpha^{\prime }\mathbf{r}^{\prime
}}(\beta-\tau).
\end{align}

Having calculated the polarizability we transform it to the reciprocal $%
\mathbf{q}$--space and boson--frequency $\nu $--representation, which
schematically is given as

\begin{align}\label{P_R_to q_def}
P_{\mathbf{rr}^{\prime }}^{\mathbf{R}^{\prime }}(\tau )\rightarrow P_{ij}^{%
\mathbf{q}}(\nu ),
\end{align}%
where indexes $i$ and $j$ refer to the product basis functions.
Transformation (\ref{P_R_to q_def}) is performed similar to the
Green's function which was specified earlier.

After that, we calculate the screened Coulomb interaction $W$. It is
convenient to divide $W$ into the bare Coulomb interaction $V$ and the
screening part $\widetilde{W}$:

\begin{equation}
W_{ij}^{\mathbf{q}}(\nu )=V_{ij}^{\mathbf{q}}+\widetilde{W}_{ij}^{\mathbf{q}%
}(\nu ).  \label{def_w_s}
\end{equation}%
In the ($\mathbf{q},\nu $)--representation we have to solve the following
linear equation system for $\widetilde{W}$:

\begin{equation}
\sum_{k}\{\delta _{ik}-\sum_{l}V_{il}^{\mathbf{q}}P_{lk}^{\mathbf{q}}(\nu )\}%
\widetilde{W}_{kj}^{\mathbf{q}}(\nu )=\sum_{k}V_{ik}^{\mathbf{q}%
}\sum_{l}P_{kl}^{\mathbf{q}}(\nu )V_{lj}^{\mathbf{q}}.  \label{W_nu}
\end{equation}%
Having found it, we switch back from $\nu -$-representation for $\widetilde{W%
}$ to $\tau $--representation and from $\mathbf{q}$ space to $\mathbf{r}$
space.

Next we find the self--energy. This is subdivided onto three steps: i) we
solve an effective Hartree--Fock band structure problem which is similar to
the familiar Hartree--Fock problem but with matrix elements of Hartree and
exchange interaction calculated using full GW Green's function from the
previous iteration. To speed up the process, we calculate exchange
interaction in real space and then transform it to the reciprocal space and
band representation. The solution of the effective Hartree--Fock problem
gives us a new exchange part of the Green's function $G^{x}$. In step ii) we
calculate the correlated part of the self energy $\Sigma ^{c}$ in ($\mathbf{r%
},\tau $)--representation. Again there are three different cases depending
on where $\mathbf{r}$ and $\mathbf{r}^{\prime }$ belong to:

\begin{align}  \label{sigc_mm3}
\Sigma^{c,\mathbf{R}^{\prime }}_{\mathbf{t}E_{1}L_{1};\mathbf{t}^{\prime
}E_{2}L_{2}}(\tau)=&-\sum_{E_{3}L_{3}}\sum_{E_{4}L_{4}}\sum_{kLk^{\prime
}L^{\prime }}  \notag \\
&\times\sum_{\alpha}\langle\varphi^{\mathbf{t}}_{E_{1}L_{1}}(\alpha)|%
\varphi^{\mathbf{t}}_{E_{3}L_{3}}(\alpha) M^{\mathbf{t}}_{kL}\rangle  \notag
\\
&\times G^{\mathbf{R}^{\prime }}_{\mathbf{t}E_{3}L_{3};\mathbf{t}^{\prime
}E_{4}L_{4}}(\tau) \widetilde{W}^{\mathbf{R}^{\prime }}_{\mathbf{t}kL;%
\mathbf{t}^{\prime }k^{\prime }L^{\prime }}(\beta-\tau)  \notag \\
&\times \sum_{\alpha^{\prime }}\langle\varphi^{\mathbf{t}^{\prime
}}_{E_{4}L_{4}}(\alpha^{\prime})|\varphi^{\mathbf{t}^{\prime}}_{E_{2}L_{2}}(\alpha^{\prime})
M^{\mathbf{t}^{\prime }}_{k^{\prime }L^{\prime }}\rangle,
\end{align}

\begin{align}  \label{sigc_mi3}
\Sigma^{c,\mathbf{R}^{\prime }}_{\mathbf{t}E_{1}L_{1};\alpha^{\prime }%
\mathbf{r}^{\prime }}(\tau)= &-\sum_{E_{2}L_{2}}\sum_{kL}\sum_{\alpha}
\langle\varphi^{\mathbf{t}}_{E_{1}L_{1}}(\alpha)|\varphi^{\mathbf{t}%
}_{E_{2}L_{2}}(\alpha) M^{\mathbf{t}}_{kL}\rangle  \notag \\
&\times G^{\mathbf{R}^{\prime }}_{\mathbf{t}E_{2}L_{2};\alpha^{\prime }%
\mathbf{r}^{\prime }}(\tau) \widetilde{W}^{\mathbf{R}^{\prime }}_{\mathbf{t}%
kL;\mathbf{r}^{\prime }}(\beta-\tau),
\end{align}

\begin{equation}
\Sigma _{\alpha \mathbf{r}\alpha ^{\prime }\mathbf{r}^{\prime }}^{c,\mathbf{R%
}^{\prime }}=-G_{\alpha \mathbf{r}\alpha ^{\prime }\mathbf{r}^{\prime }}^{%
\mathbf{R}^{\prime }}(\tau )\widetilde{W}_{\mathbf{r}\mathbf{r}^{\prime }}^{%
\mathbf{R}^{\prime }}(\beta -\tau ),  \label{sigc_ii3}
\end{equation}%
In iii) we transform the self--energy back to the band representation in $%
\mathbf{k}$--space, using the formulae similar to Green's function. We also
transform it from $\tau $ to Matsubara's $\omega $--frequency.

The last part is to solve Dyson's equation in order to find new correlated
part of the Green's function $G^{c}$. We perform this step using band
representation in $\mathbf{k}$--space:

\begin{align}
& \sum_{\lambda ^{\prime \prime }}\{\delta _{\lambda \lambda ^{\prime \prime
}}-G_{\lambda }^{x}(\mathbf{k};\omega )\Sigma _{\lambda \lambda ^{\prime
\prime }}^{c}(\mathbf{k};\omega )\}G_{\lambda ^{\prime \prime }\lambda
^{\prime }}^{c}(\mathbf{k};\omega )  \notag  \label{G_k_omega_lin} \\
& =G_{\lambda }^{x}(\mathbf{k};\omega )\Sigma _{\lambda \lambda ^{\prime
}}^{c}(\mathbf{k};\omega )G_{\lambda ^{\prime }}^{x}(\mathbf{k};\omega ).
\end{align}%
This is accompanied by finding new chemical potential $\mu $, with
total--electron--number--conservation condition. Then, we transform Green's
function back from ($\mathbf{k},\omega $)-- to ($\mathbf{k},\tau $%
)--representation, and use it to calculate new electronic density and new
Hartree potential which are needed for the next iteration. This closes our
iteration cycle. It is important to mention that we completely avoid
convolutions in $\mathbf{k}$--space, which saves a lot of computer time as
compared to pure $\mathbf{k}$--space implementation.

We can also perform quasiparticle self--consistent calculations. Different
from the QPscGW method by Kotani et al.\cite{prb_76_165106}), our method is
based exclusively on imaginary axis data: We approximate frequency
dependence of the self--energy by a linear function near zero Matsubara's
frequency and reduce the problem to the solution of Dyson's equation to one
matrix diagonalization. Then, as justified in [Ref.\onlinecite{prb_76_165106}%
] we neglect by Z--renormalization of Green's function  and use it as an
input for the new self--consistent iteration.

To get single--particle densities of states (DOS) from the full
self--consistent GW approximation we perform similar linear approximation to
the self--energy and compute spectra as the final step after the
self--consistency is reached. For the low energy behavior of the spectral
functions this kind of analytical continuation is a lot more stable and
produces essentially the same DOS as the traditional Pade approximation.

\section{Details of Calculations}

\label{details}

Parameters of our calculations are as follows: We use mesh $7\times 7\times 7
$ in the Brillouin zone. Green's function was expanded over Bloch states
obtained from LDA\ based full potential LAPW band structures. The number of
bands in this expansion varies between 142 and 168 depending on the $\mathbf{%
k}$--point in the Brillouin zone. Note that such large number of states is
only possible only when using real--space based implementation of the GW
method while  using reciprocal space, it is very hard to handle more than
40--50 bands in the LAPW based GW method.

Inside the MT spheres we expand the functions of fermionic type (Green's
function and self--energy) in spherical harmonics up to $l_{max}=5$. Bosonic
functions (polarizability and interaction) are expanded up to $l_{max}=6$.
In the interstitial region each function is expanded in plane waves. We use
more plane waves for bosonic functions (250--300) than for fermionic ones.
Our full basis size to expand bosonic functions both inside the MT spheres
and in the interstitials is about 600 depending on the particular $\mathbf{k}
$--point.

All calculations are performed for the temperature 1000K. The LDA
calculations use exchange--correlation parametrization after Perdew
and Wang.\cite{prb_45_13244}

\section{Results}

\label{Res}

\begin{table}[tbp]
\caption{$5f$ occupation numbers for $\protect\delta $--Pu (taken at
the volume of its $\protect\delta $--phase) obtained within scalar
relativistic (SR) and fully relativstic (FR)\ approaches.}
\begin{center}
\begin{tabular}{@{}cccc}
Method & $5f_{5/2}$ & $5f_{7/2}$ & $5f_{5/2}+5f_{7/2}$ \\ \hline
LDA, SR &  &  & 5.17 \\
LDA, FR & 4.15 & 0.92 & 5.07 \\
GW, SR &  &  & 4.82 \\
QP, FR & 4.26 & 0.56 & 4.84 \\
GW, FR & 4.45 & 0.44 & 4.89%
\end{tabular}%
\end{center}
\label{occup_5f_delta}
\end{table}

\begin{table}[tbp]
\caption{$5f$ occupation numbers for $\protect\delta $--Pu (taken at the
volume of its $\protect\alpha $--phase) obtained using  fully relativstic
(FR)\ approach.}
\begin{center}
\begin{tabular}{@{}cccc}
Method & $5f_{5/2}$ & $5f_{7/2}$ & $5f_{5/2}+5f_{7/2}$ \\ \hline
LDA, FR & 3.72 & 1.38 & 5.10 \\
GW, FR & 4.05 & 0.81 & 4.86%
\end{tabular}%
\end{center}
\label{occup_5f_alpha}
\end{table}

\begin{table}[tbp]
\caption{$5f$ occupation numbers for fcc--Americium obtained using fully
relativstic (FR)\ approach.}
\begin{center}
\begin{tabular}{@{}cccc}
Method & $5f_{5/2}$ & $5f_{7/2}$ & $5f_{5/2}+5f_{7/2}$ \\ \hline
LDA, FR & 5.36 & 0.84 & 6.2 \\
QP, FR & 5.66 & 0.26 & 5.92 \\
GW, FR & 5.67 & 0.27 & 5.94%
\end{tabular}%
\end{center}
\label{occup_5f_am}
\end{table}

\begin{figure}[t]
\centering
\rotatebox{-90}{
\includegraphics[width=6.0 cm]{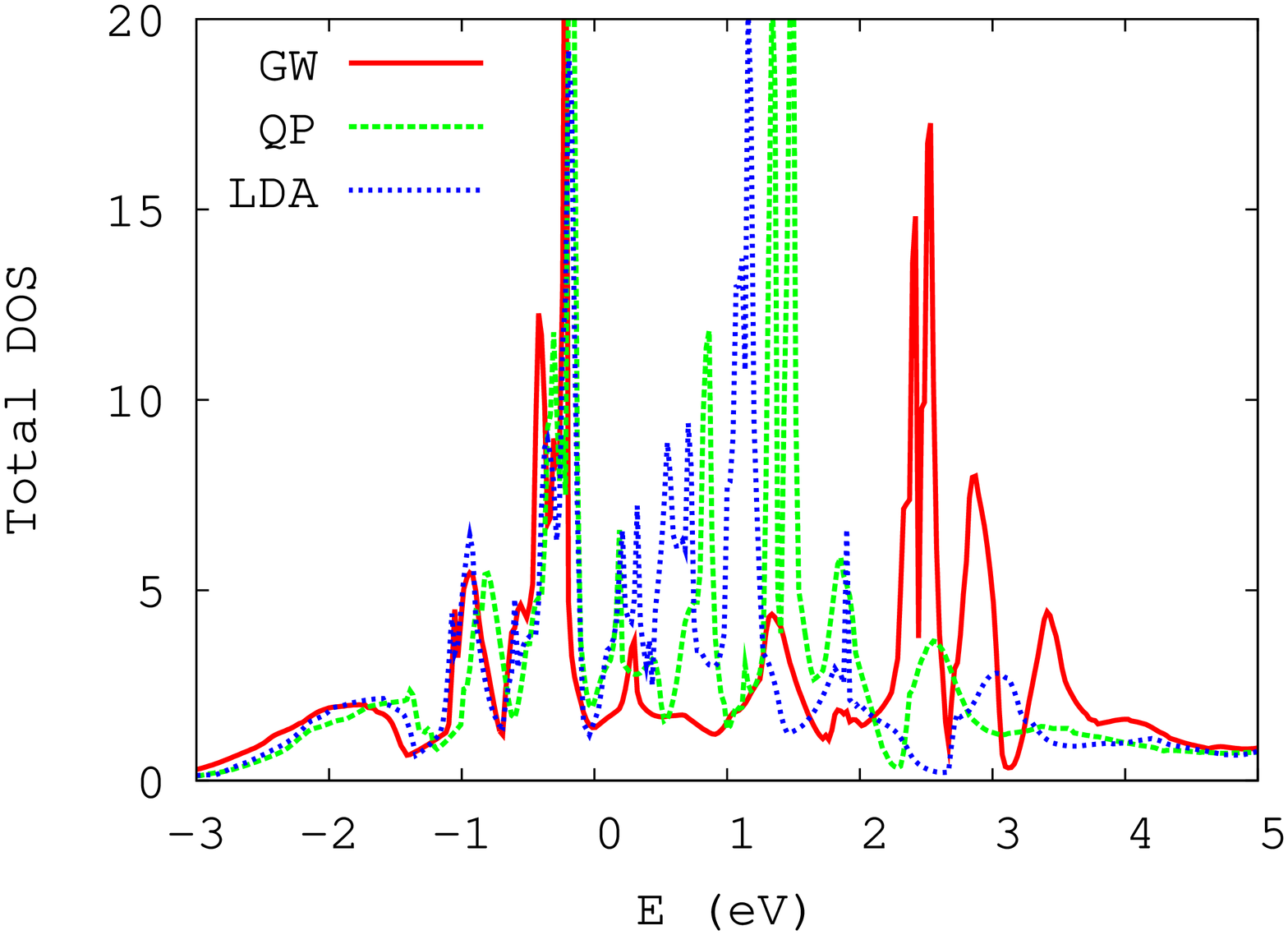} }
\caption{(Color online) Total density of states (DOS) of $\protect\delta $%
--Plutonium as obtained in self consistent relativistic calculations.
Comparison is made between GW, QP, and LDA approaches.}
\label{dos_rgw_qp_lda_delta}
\end{figure}

\begin{figure}[t]
\centering
\rotatebox{-90}{
\includegraphics[width=6.0 cm]{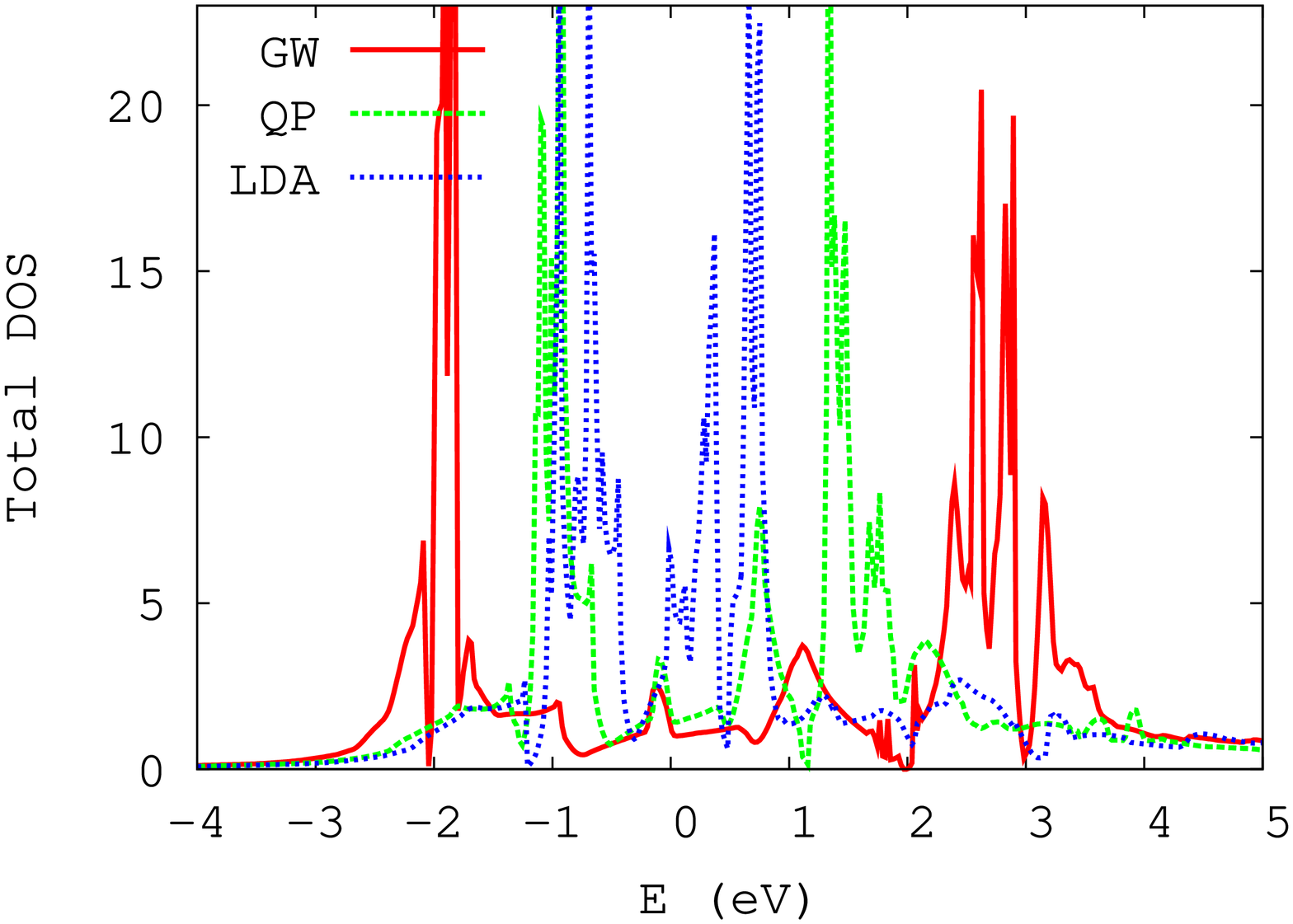} }
\caption{(Color online) Total density of states (DOS) of Americium as
obtained in self consistent relativistic calculations. Comparison is made
between GW, QP, and LDA approaches.}
\label{dos_rgw_qp_lda_am}
\end{figure}

\begin{figure}[b]
\centering
\rotatebox{-90}{
\includegraphics[width=6.0 cm]{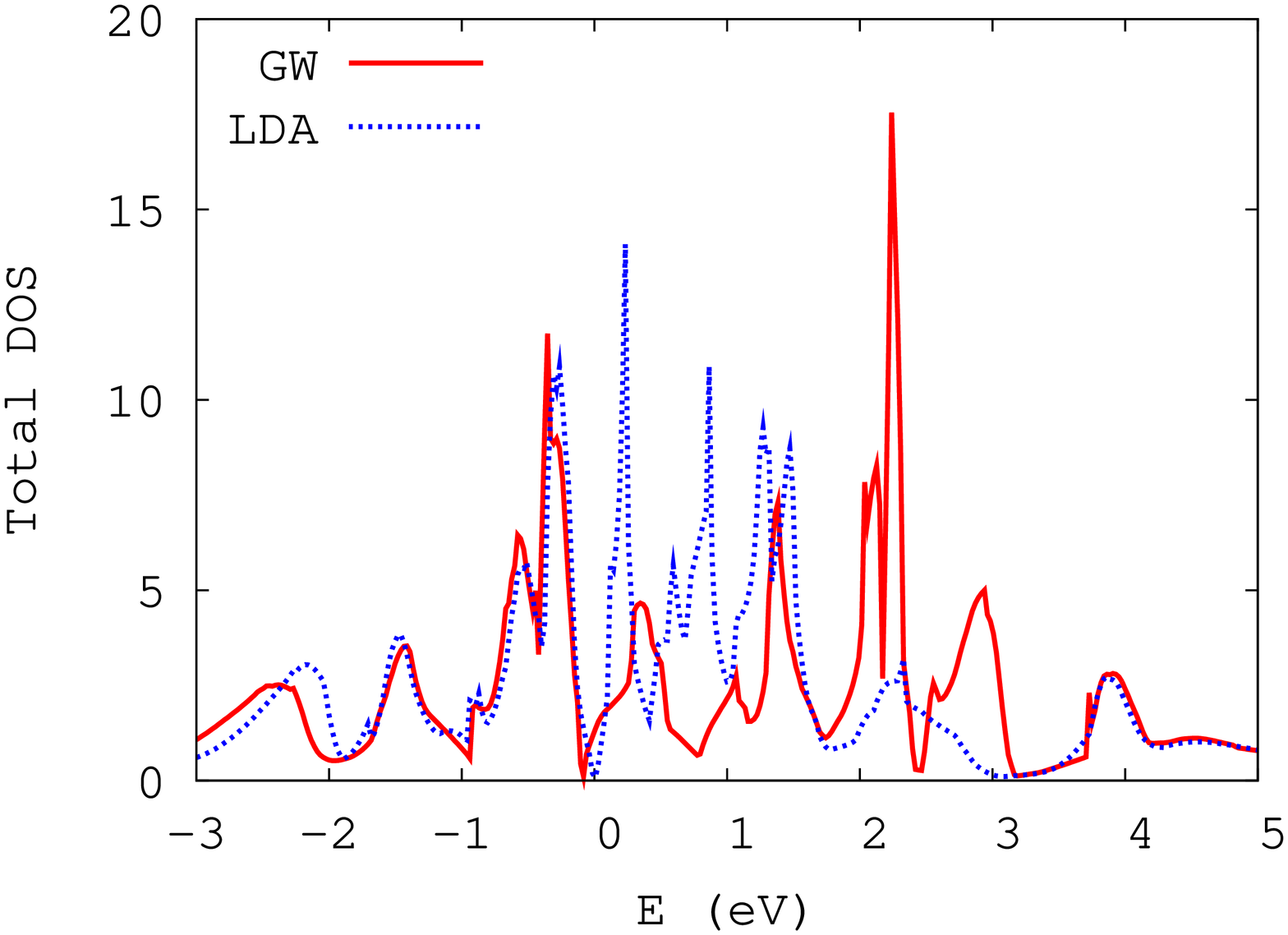} }
\caption{(Color online) Total density of states (DOS) of Plutonium
taken at the volume of its $\protect\alpha $--phase as obtained in
self consistent relativistic calculations. Comparison is made
between GW and LDA approaches.} \label{dos_rgw_lda_alpha}
\end{figure}

We first discuss our results obtained by various methods for the number of $%
5f$ electrons, $n_{5f},$ as given in Tables \ref{occup_5f_delta}, \ref%
{occup_5f_alpha}, and \ref{occup_5f_am}. As follows from experiment\cite%
{prl_90_196404,prl_93_097401,prb_72_085109,prb_76_073105}, the $5f$
occupation in Pu is close to 5, and the corresponding occupation in Am is
close to 6. Our scalar--relativistic GW result (4.82) is very close to the
value 4.85 obtained in the calculation performed by Chantis et al.\cite%
{philmag_89_1801} As it is seen from the calculated data our GW results are
consistently less than the experimental ones, which may be attributed in
part to the fact that we count $5f$ electrons only inside the MT spheres. In
this respect, the LDA results, which are a little too large, look less
consistent with experiment. There is also a noticeable difference between
LDA and GW in the separation of $n_{5f}$ onto $5f_{5/2}$ and $5f_{7/2}$
contributions where the GW approximation produces more $5f_{5/2}$ electrons
and less $5f_{7/2}$ electrons. An interesting trend is seen when one looks
at the volume dependence of $5f$ counts for Plutonium (Tables \ref%
{occup_5f_delta} and \ref{occup_5f_alpha}). Full $5f$ occupation is
amazingly unchanged but the distribution between $5f_{5/2}$ and $5f_{7/2}$
states changes a lot.

\begin{figure}[t]
\centering
\rotatebox{-90}{
\includegraphics[width=6.0 cm]{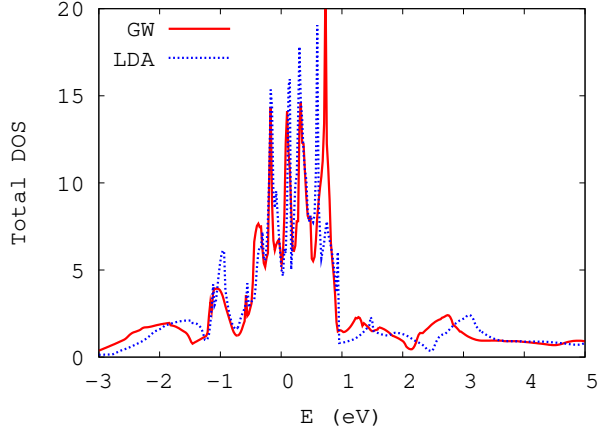} }
\caption{(Color online) Total density of states (DOS) of $\protect\delta $%
--Plutonium as obtained in self--consistent scalar--relativistic
calculations. Comparison is made between GW and LDA approaches.}
\label{dos_srgw_lda_pu}
\end{figure}

\begin{figure}[t]
\centering
\rotatebox{-90}{
\includegraphics[width=6.0 cm]{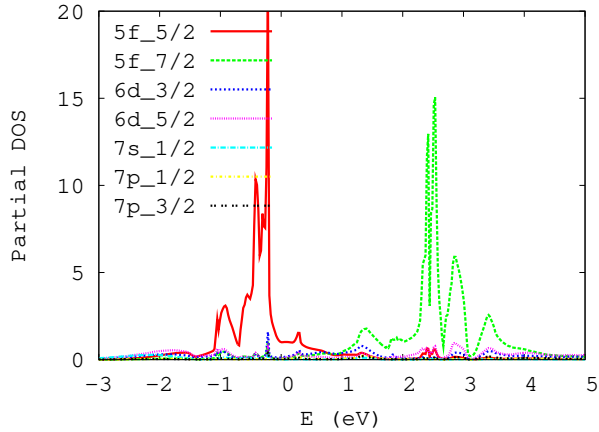} }
\caption{(Color online) Partial densities of states (PDOS) for
Plutonium  (taken at the volume of its $\protect\delta $--phase) as
obtained in self--consistent relativstic GW calculation.}
\label{pdos_rgw_delta}
\end{figure}

\begin{figure}[t]
\centering
\rotatebox{-90}{
\includegraphics[width=6.0 cm]{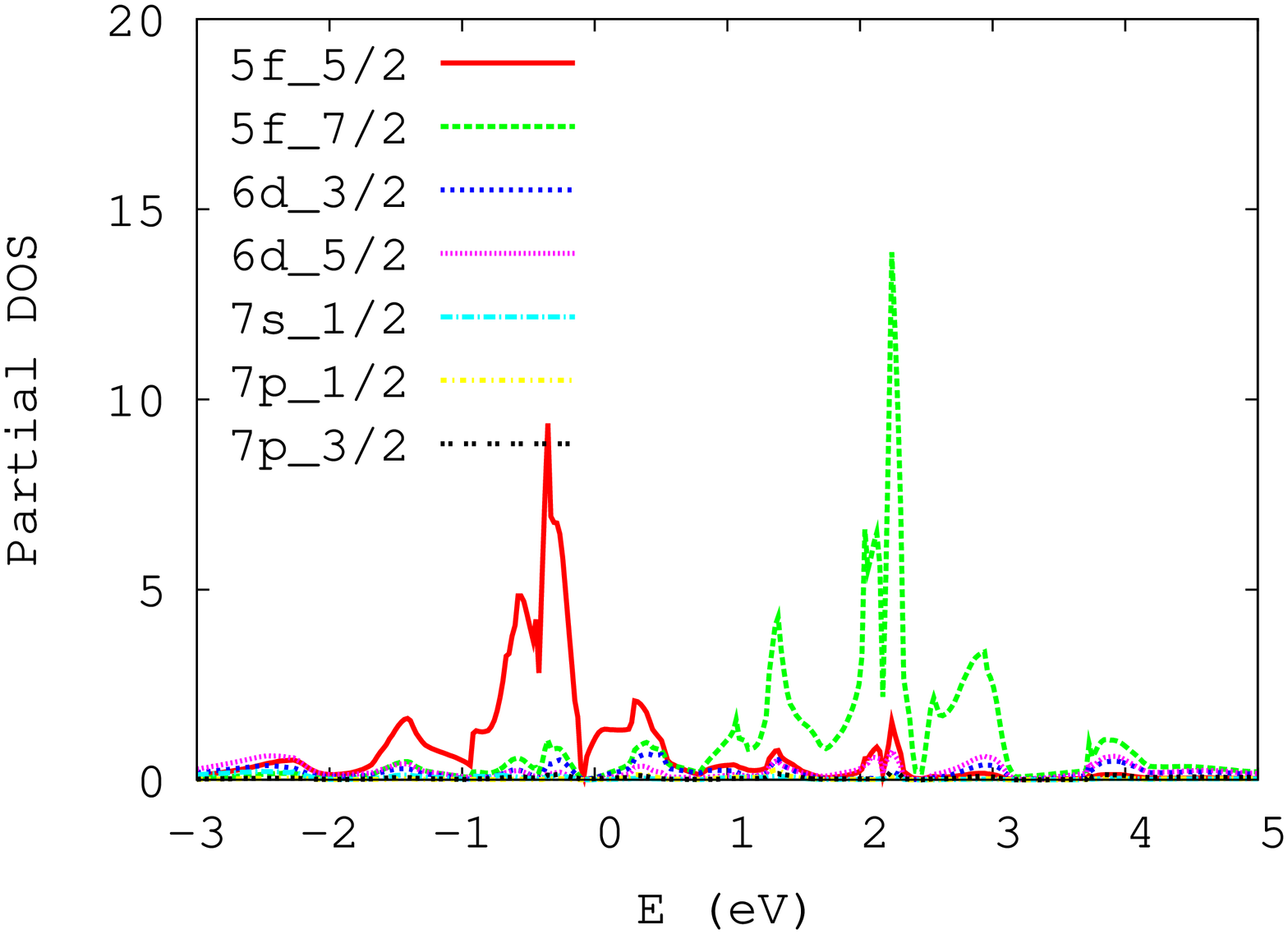} }
\caption{(Color online) Partial densities of states (PDOS) for
Plutonium (taken at the volume of its $\protect\alpha $--phase) as
obtained in self--consistent relativstic GW calculation.}
\label{pdos_rgw_alpha}
\end{figure}

\begin{figure}[tb]
\centering
\rotatebox{-90}{
\includegraphics[width=6.0 cm]{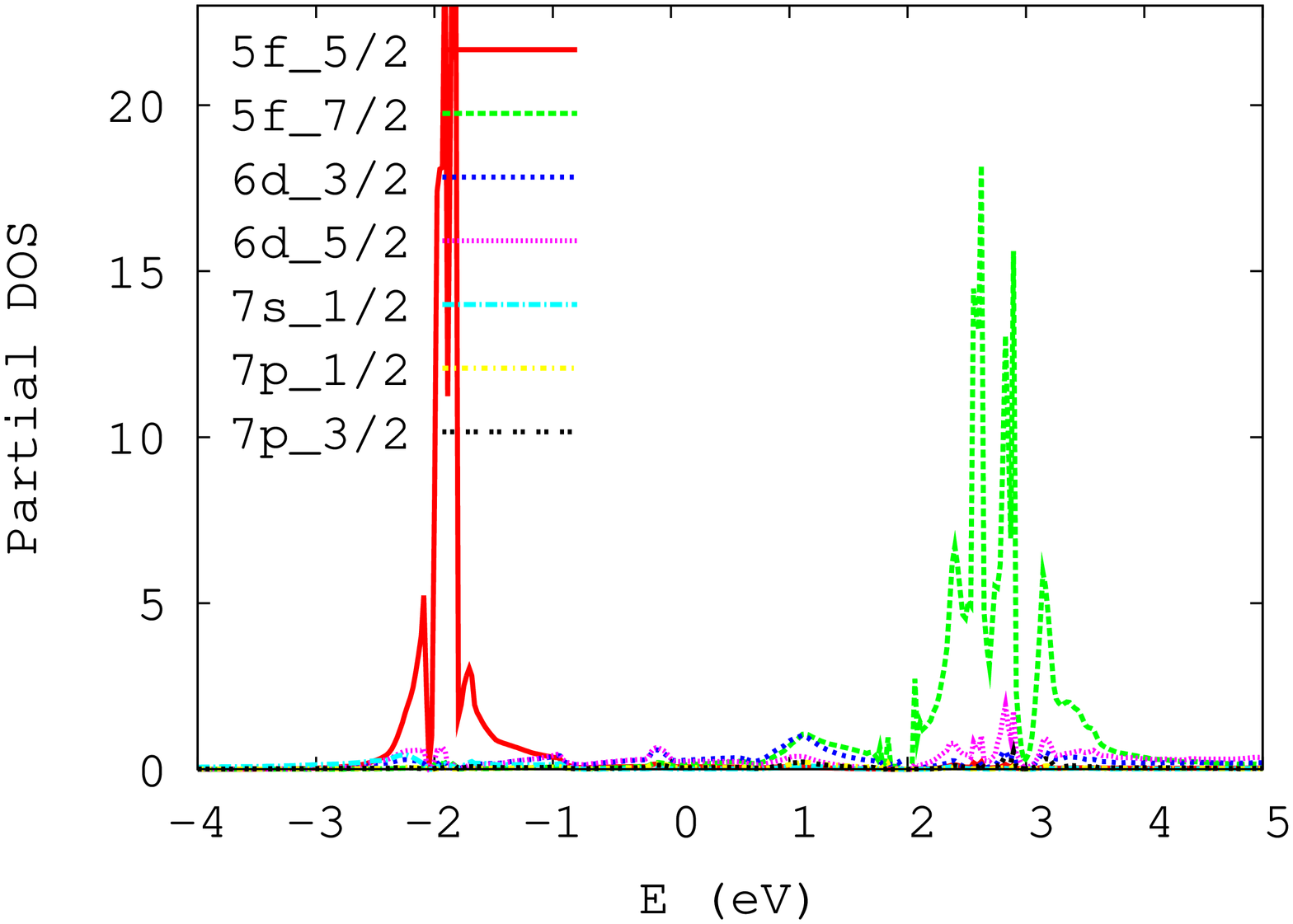} }
\caption{(Color online) Partial densities of states (PDOS) for Americium as
obtained in self--consistent relativstic GW calculation}
\label{pdos_rgw_am}
\end{figure}

\begin{figure}[tb]
\centering \rotatebox{-90}{
\includegraphics[width=6.0 cm]{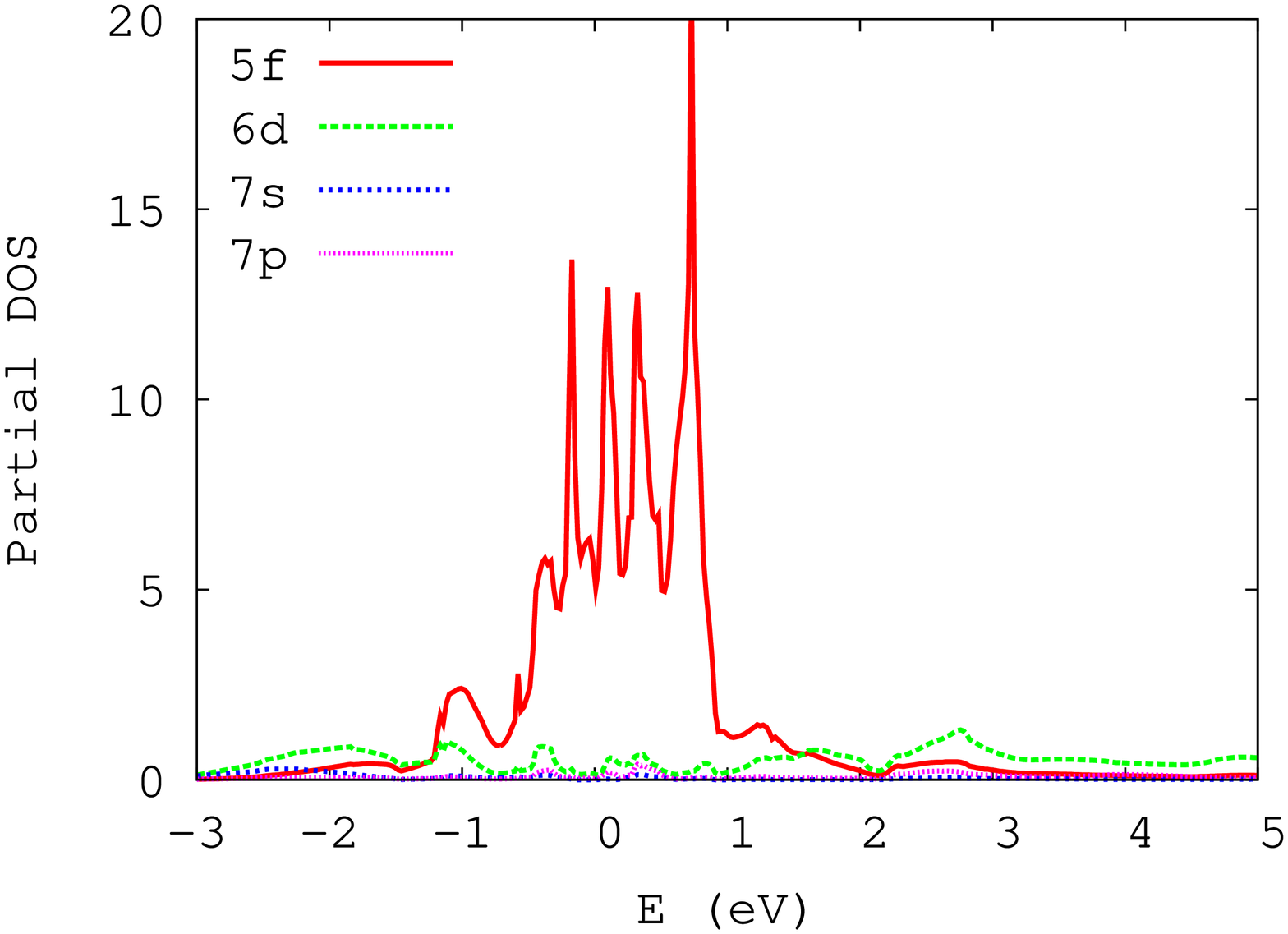} }
\caption{(Color online) Partial densities of states (PDOS) for $\protect%
\delta $--Plutonium as obtained in self--consistent scalar--relativistic GW
calculation.}
\label{pdos_srgw_delta}
\end{figure}

We next describe our calculated total densities of states (DOS) (Figures \ref%
{dos_rgw_qp_lda_delta}-\ref{dos_srgw_lda_pu}) and partial densities of
states (PDOS)(Figures \ref{pdos_rgw_delta}-\ref{pdos_srgw_delta}). In all
plots chemical potential is set to zero. For $\delta $--Pu (Fig.\ref%
{dos_rgw_qp_lda_delta}) we notice that the occupied part of the spectrum as
obtained using GW, LDA, or QP is practically indistinguishable while the
unoccupied part is different. Here, the GW method spreads the spectrum over
a wide energy interval while the LDA produces features mostly near the Fermi
level.

The spectrum of Americium (Fig.\ref{dos_rgw_qp_lda_am}) shows no
similarity between different methods even for the occupied part of
the DOS. Here we clearly see the advantage of using the GW method
which gives the lowest position of the peak at minus 2~eV in much
better agreement with the experimental value\cite{prl_52_1834}
(minus 2.8~eV) than the LDA or QP approaches do. The unoccupied part
of the spectrum gets progressively wider when we go from LDA to QP
and then to full GW calculation.

Calculated electronic structure of $\delta $--Pu at a reduced volume,
corresponding to the volume of $\alpha $--phase (Fig.\ref{dos_rgw_lda_alpha}%
) in general shows broader features than the one obtained for the $\delta $%
--Pu volume. The difference between LDA and GW calculations seems to be
reduced.

The DOS of $\delta $--Pu as obtained in scalar--relativistic calculation
(Fig.\ref{dos_srgw_lda_pu}) differs on a qualitative level from the
relativistic result, therefore,  it is quite clear that any serious
calculation for this element should take into account spin--orbit
interaction.

PDOS from all calculations performed in our work tell us that $5f$ states in
Pu and Am play a key role in energy region close to the Fermi level. We see
the increase in hybridization between $5f_{5/2}$ and $5f_{7/2}$ states when
we go from $\delta $--Pu (Fig.\ref{pdos_rgw_delta}) to $\alpha $--Pu (Fig.%
\ref{pdos_rgw_alpha}), and we see practically perfect separation between
these states in Americium metal (Fig.\ref{pdos_rgw_am}).

\begin{figure}[t]
\centering \rotatebox{-90}{
\includegraphics[width=6.0 cm]{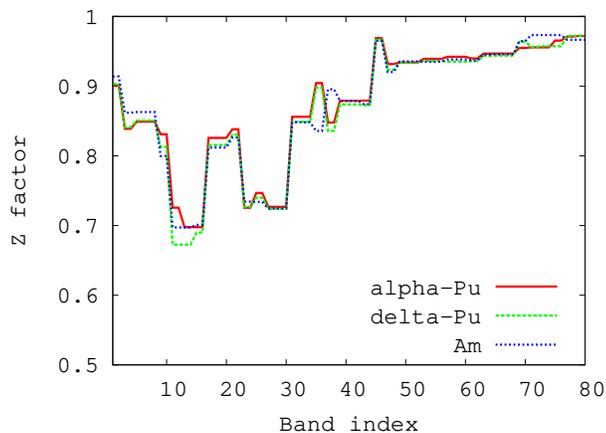} }
\caption{(Color online) Band renormalization factor Z  as a function of band
index for fcc--Plutonium (taken at volumes of $\protect\alpha $%
-- and $\protect\delta $--phases) and for fcc--Americium as obtained in
self--consistent relativistic GW calculations for $\mathbf{k}=(0,0,0)$.}
\label{z_rgw}
\end{figure}

\begin{figure}[tb]
\centering \rotatebox{-90}{
\includegraphics[width=6.0 cm]{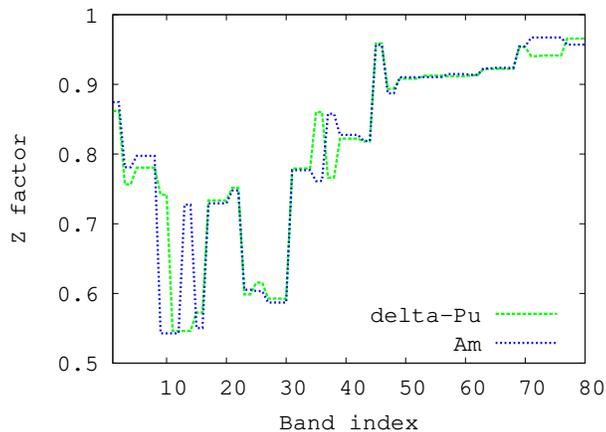} }
\caption{(Color online) Band renormalization factor Z as a function of band
index for fcc--Plutonium (taken at volume of the $\protect%
\delta $--phase) and for fcc--Americium as obtained in
self--consistent relativistic QP calculations for
$\mathbf{k}=(0,0,0)$.} \label{z_rqp}
\end{figure}

The difference between QP and self--consistent GW electronic structures
becomes more clear when we consider the quasiparticle renormalization factor
$Z$ (Fig.\ref{z_rgw} and \ref{z_rqp}). We calculate $Z$ factor in band
representation according to

\begin{align}  \label{z_def}
Z^{\mathbf{k}}_{\lambda\lambda^{\prime }}=\big(\mathbf{1}-\frac{\partial
\mathbf{\Sigma}^{\mathbf{k}}(\omega)}{\partial \omega}\big)%
^{-1}_{\lambda\lambda^{\prime }}\big|_{\omega\rightarrow 0}.
\end{align}

In our case, the indexes ($\lambda ,\lambda ^{\prime }$) correspond to the
effective Hartree--Fock band structure problem where Hartree and exchange
interactions are calculated using full GW Green's function. The more Z
differs from 1 the stronger DOS differs from the effective Hartree--Fock\
band structure which usually has too broad spectral features. On Fig.\ref%
{z_rgw} and \ref{z_rqp} we have plotted the diagonal components of Z factor
matrices as functions of the band index for 80 lowest bands for the $\mathbf{%
k}=(0,0,0)$ point of the Brillouin zone. In all cases the position of the
Fermi level is between band 16 and band 17. Actually there are 6
distinguishable bands ($5f_{5/2}$) below $E_{f}$ and 8 bands ($5f_{7/2}$)
above $E_{f}$ which have noticeably smaller Z's than the rest of the
spectrum. It is also clearly seen that Z's for the $f$--bands in the QP
calculation ($0.55\div 0.6$) are smaller than those obtained in the
self--consistent GW calculation ($0.65\div 0.75$). This explains why
spectral features in the QP electronic structure are closer to the $E_{f}$.
One can say that in case of Am and Pu the QP approximation looks like being
overscreened similar to the LDA.

In conclusion, we have described our implementation of the relativistic
self--consistent GW method and its application to the electronic structure
for Plutonium and Americium metals. We have found that the inclusion of
relativistic effects\ in GW is extremely important for the proper treatment
of the actinides. We also discussed the differrences in spectral functions
obtained using the present approach with LDA and quasiparticle
self--consistent GW approximations.

\section*{ACKNOWLEDGEMENTS}

\label{ack}

This work was supported by the United States Department of Energy Nuclear
Energy University Program, Contract No. 00088708. We would like to thank
V.~Oudovenko for adaptating our GW code to a computer cluster at Rutgers
University.


\end{document}